# High-Pressure Evolution of the Specific Heat of a Strongly Underdoped Ba(Fe$_{0.963}$Co$_{0.037}$)As$_2$ Iron-Based Superconductor

Y. Zheng[1], Y. Wang[1], F. Hardy[2], A. E. Böhmer[2,3], T. Wolf[2], C. Meingast[2], and R. Lortz[1*]

[1]*Department of Physics, Hong Kong University of Science & Technology, Clear Water Bay, Kowloon, Hong Kong*
[2]*Karlsruhe Institute for Technology, IFP, PO Box 3640, 76021 Karlsruhe, Germany*
[3]*Fakultät für Physik, Karlsruhe Institute of Technology, 76128 Karlsruhe, Germany*

We report specific-heat experiments under the influence of high pressure on a strongly underdoped Co-substituted BaFe$_2$As$_2$ single crystal. This allows us to study the phase diagram of this iron pnictide superconductor with a bulk thermodynamic method and pressure as a clean control parameter. The data show large specific-heat anomalies at the superconducting transition temperature, which proves the bulk nature of pressure-induced superconductivity. The transitions in the specific heat are sharper than in resistivity, which demonstrates the necessity of employing bulk thermodynamic methods to explore the exact phase diagram of pressure-induced Fe-based superconductors. The $T_c$ at optimal pressure and the superconducting condensation energy are found to be larger than in optimally Co-doped samples at ambient pressure, which we attribute to a weak pair breaking effect of the Co ions.

## I. INTRODUCTION

The AEFe$_2$As$_2$ (AE = Ba, Sr, Ca), or '122' compounds remain one of the most intensively studied families of the iron-based pnictide superconductors. Comparatively large high-quality single crystals are available. Their phase diagram can be widely explored upon introduction of hole or electron dopants, or by application of external hydrostatic [1-5] or internal chemical pressure [6]. For example, substitution of Ba$^{2+}$ by K$^+$ introduces holes [7]. Substitution of Fe$^{2+}$ by Co$^{3+}$ [8] introduces electrons. Isovalent doping can also introduce superconductivity, for example upon substitution of Fe by Ru [9] or As by P [10], where the effect of the introduced smaller ions can be regarded as an internal chemical pressure effect. The phase diagram shows many similarities to numerous heavy-fermion [11] and organic superconductors [12]: a magnetic transition gets gradually suppressed as a function of a control parameter, with a superconducting phase developing around the point where this transition extrapolates to zero temperature. In Ba122, the magnetic transition is of an antiferromagnetic (AFM) spin density wave (SDW) nature which nearly coincides with a structural transition from a high-temperature tetragonal (T) to a low-temperature orthorhombic (O) structure [13]. The close vicinity of the two transition lines demonstrates a strong coupling between magnetism and crystalline structure [14-18]. Various experiments [19-27] showed that in Co-substituted Ba122, the AFM phase coexists homogeneously with superconductivity. It has been proposed that the exact phase diagram in the coexistence region may serve as a test for the symmetry of the Cooper-pair wave function [28] and the most likely candidate is the s+- superconducting state [13,17,28]. A detailed knowledge of the phase transitions in this region is therefore of particular importance to gain new insights into the mechanism of superconductivity in the pnictides.

---

[*] Corresponding author: lortz@ust.hk

The specific heat of Co-doped Ba122 has been intensively studied on a set of single crystals of various Co-contents [29]. One drawback of this approach is that many different samples are used in such an investigation and disorder related to the non-stoichiometry of the Co-substitution may have significant impacts on the phase diagram. Pressure experiments of Co substituted Ba122 samples indicated that Co substitution has a very similar effect to the application of hydrostatic pressure [18,30,31] and the phase diagram of Ba122 can be traced by both methods, although the exact doping mechanism remains unclear [31,32]. Through application of pressure the phase diagram can be investigated starting from one single-crystalline sample of fixed composition. Pressure represents therefore a particularly clean control parameter as the influence of crystalline disorder throughout the study is held constant, and in this paper we will follow this approach. Superconductivity has been induced previously through application of high pressure in the parent compound Ba122, but to the best of our knowledge bulk thermodynamic methods under pressure have not been reported. In the present experiment, we start from a strongly underdoped $Ba(Fe_{0.963}Co_{0.037})As_2$ single crystal at the borderline of superconductivity, with the aim of investigating whether Co-doping and pressure have a similar impact on the superconducting transition. The method we are using is a rarely performed AC specific-heat technique [33,34] in a Bridgman type of pressure cell in combination with concurrent resistivity measurements.

## II. EXPERIMENTAL

Single crystals were grown from self-flux using pre-reacted FeAs and CoAs powders mixed with Ba in glassy carbon crucibles. The crucible was sealed in an evacuated $SiO_2$ ampoule and heated to 650 ° C and then to 1200 ° C with holding times of 5 h. The growth took place upon subsequent cooling at a rate of 1 °C / h. At 1000 ° C, the ampoule was tilted to decant the remaining liquid flux from the crystals and then removed from the furnace. The composition of these crystals was determined by energy dispersive x-ray spectroscopy and 4-circle diffractometry. The Co-doping of the crystal was carefully chosen to be just at the underdoped border of superconductivity. Figure 1 shows specific heat data at ambient pressure of 3 single crystals with 3.5, 3.7 and 4 % of the Fe atoms replaced by Co, measured in our micro-relaxation calorimeter. The specific heat of the 3.5 % sample shows no sign of superconductivity, while the 4 % sample shows a clear superconducting transition at 6.2 K. The intermediate 3.7 % sample (chosen for our high pressure study in this paper) shows only a tiny but sharp superconducting jump at $T$ = 6.7 K, which becomes visible only after subtraction of the phonon background. This indicates that the superconductivity in this 3.7 % Co-doped sample is still of filamentary nature, while it becomes bulk superconductivity for marginally stronger doping.

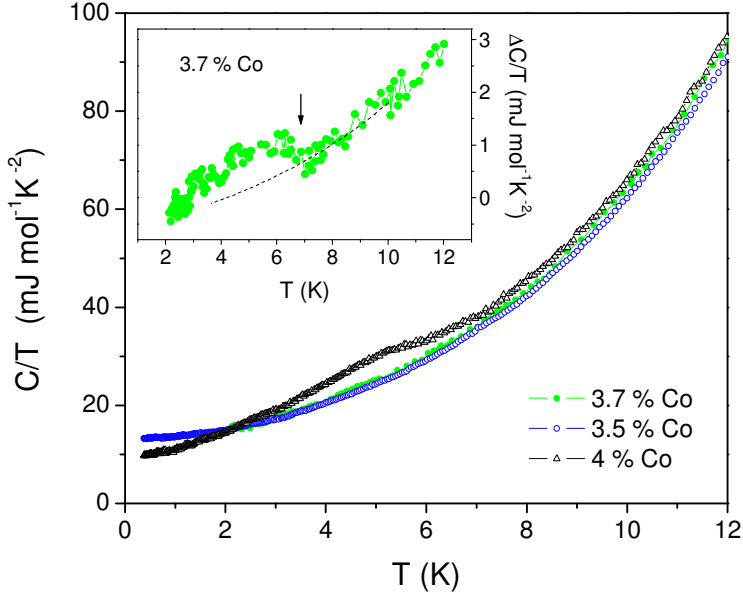

**Figure. 1** Specific heat of BaFe$_2$As$_2$ single crystals with 3.5, 3.7 and 4 % of the Fe atoms replaced by Co. The inset shows the tiny superconducting transition in the specific heat of Ba(Fe$_{0.963}$Co$_{0.037}$)As$_2$ after subtraction of the normal state background, which has been obtained from the data of the non-superconducting Ba(Fe$_{0.965}$Co$_{0.035}$)As$_2$.

Figure 2 shows a photo of the experiment mounted in a pyrophillite (Al$_2$Si$_4$O$_{10}$(OH)$_2$) gasket on a tungsten-carbide anvil of our Bridgman cell. The white background represents a disk of steatite. 12 Au leads enter the cell through thin grooves which were filled later with compressed pyrophillite powder. The junctions of two thermocouples were placed on the sample with their ends heat-sinked to the Au leads at the edge. To ensure optimal performance, we combined a type E thermocouple with a Chromel/AuFe(0.07%) thermocouple. The contact resistance of one thermocouple served as a resistive Joule heater, while with the second one an induced temperature modulation was monitored. Dependent on the temperature range and as a check of consistency, their roles could be exchanged. Additional Au terminals served for concurrent resistivity measurements. The thin silver-colored strip on the lower left side is a thin Pb foil in a 4-wire electrical configuration. The superconducting transition of Pb is pressure dependent and, with the help of literature data [35-38], serves as a sensitive manometer. After completing the set-up, a second steatite disk had been placed on top and the gasket was pressurized between two anvils in the cylindrical body of the pressure cell. The soft 'soap stone' steatite served as pressure medium. It offers quasi-hydrostatic conditions, with the advantage that it is solid from the beginning thus avoiding additional shear stress upon cooling through the solidification transition of a liquid medium. Most importantly, its comparatively low thermal conductance facilitates achieving a thermal isolation of the sample from the anvils. Specific-heat experiments under high-pressure conditions are extremely difficult, especially at elevated temperatures beyond ~10 K where the thermal conductance of the pressure media increases rapidly. Up to now, we succeeded in this temperature range only with steatite. Its drawback is certainly that our samples are exposed to some pressure gradients, which are known to have a rather strong impact on the phase diagram of Ba122 [3,18]. However, with special care taken to ensure a perfectly parallel anvil alignment, and adequate waiting time (2-3 days) for the cell to relax after each pressure change, our Pb manometer showed sharp superconducting transitions of width $\Delta T_c \leq 0.05$ K up to the maximum pressure. This shows that pressure gradients did not exceed 2 kbar, which is comparable to what has been reported for some liquid pressure media [3].

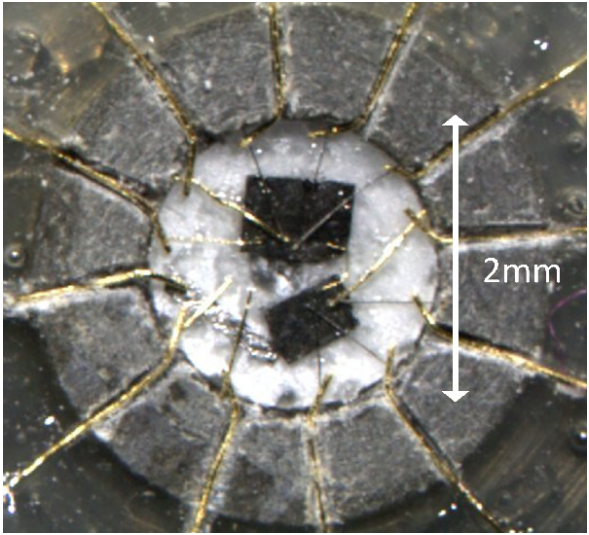

**Figure. 2** Photograph of the experiment mounted in a pyrophyllite gasket (dark-gray ring with gold wires as electrical feedthroughs) of a Bridgman pressure cell with a solid pressure transmitting medium. The Ba(Fe$_{0.963}$Co$_{0.037}$)As$_2$ Sample under investigation is represented by the lower black rectangular shaped objects. A second sample represented by the upper black rectangular object is for a separate experiment, not related to this study.

The high-pressure heat-capacity experiments were carried out with an AC alternating-temperature technique. A standard model of AC calorimetry [39] relates amplitude and phase shift of the induced temperature modulation $T_{AC}$ to the heat capacity ($C_p$) of the sample and the thermal conductance ($K$) of the surrounding pressure medium: $T_{AC} = P_0 / [K + i\omega C_p]$ ($P_0$ is the heating power). If the frequency $\omega$ of the temperature modulation is chosen sufficiently high (200 Hz – 1 kHz), the heat capacity term dominates and the thermal conductance can be neglected to a good approximation. Owing to the difficulty to exactly model the heat flow through the cell, the information on the absolute value of the specific heat is limited and the data are presented in arbitrary units. Nevertheless, the method represents a powerful high-resolution technique for studying the pressure evolution of thermodynamic phase transitions. $P_0$ has been monitored carefully in order to compare the data at different pressures on the same scale. The electrical resistivity was measured with a Keithley$^{TM}$ 6221 AC-current source in combination with a digital lock-in amplifier. The frequency was chosen as a few Hz in order to minimize phase shifts due to dissipation or capacitive effects.

# III. RESULTS

Figure 3 shows resistivity data at ambient pressure and at various pressures up to 4.4 GPa. At ambient pressure, the resistivity goes through a shallow minimum around 120 K, which can be attributed to the T-O transition. Below 21 K, the value decreases gradually, but the total variation to the lowest temperature represents only 15 % of the normal-state value at 21 K. It is approaching a temperature-independent value at 6 K at the temperature where the specific heat shows the tiny jump-like anomaly, which we attribute to filamentary superconductivity. At 0.3 GPa, the minimum occurs at slightly lower temperature (96 K) and the overall trend becomes of more metallic nature. The superconducting transition remains very broad, which is likely related to stress under non-hydrostatic conditions: Our pressure cell achieves optimal quasi-hydrostatic conditions only above ~ 1 GPa, which (together with the strong pressure dependence of $T_c$ in this pressure range) explains the broadness. A small reversible step occurs at 27 K that will be discussed later. At higher pressure, the minimum disappears and the superconducting transition becomes sharp, indicating that the large width of the superconducting transition in 0.3 GPa is not caused by poor sample quality. At the maximum pressure, the temperature dependence of the resistivity above $T_c$ is perfectly linear, as well-known from optimally doped cuprate superconductors [40] and optimally-doped Ba122 compounds [41], and generally attributed to a non-Fermi liquid behavior in the vicinity of optimal doping [42].

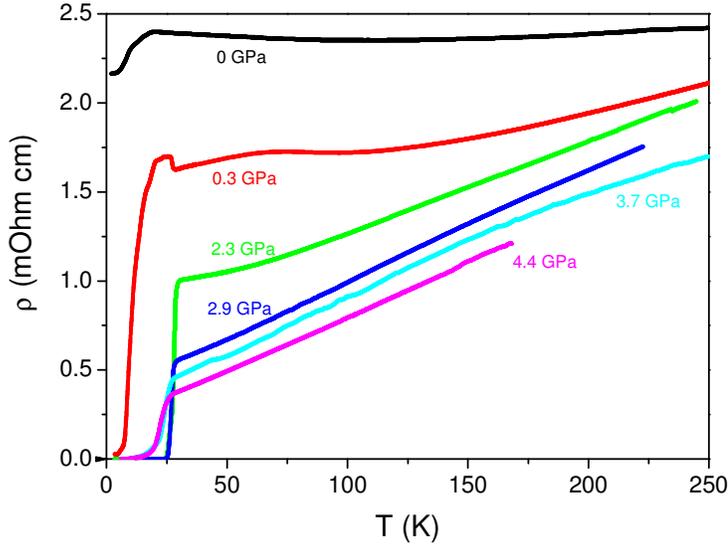

**Figure. 3** Resistivity data of a Ba(Fe$_{0.963}$Co$_{0.037}$)As$_2$ single crystal at ambient pressure and under various pressures up to 4.4 GPa.

Figure 4 shows the specific heat at various pressures between 0.3 and 4.4 GPa, together with the resistivity data. In (f) we present the total specific heat, and in panel (a) to (e) we subtracted an approximate phonon background to show the phase transition anomalies in comparison with the resistivity data for each pressure more clearly. All data has been taken sequentially upon increasing pressure. At the lowest pressure (0.3 GPa), a comparatively small and broad jump-like superconducting transition is visible in the specific heat (Figure 4a).

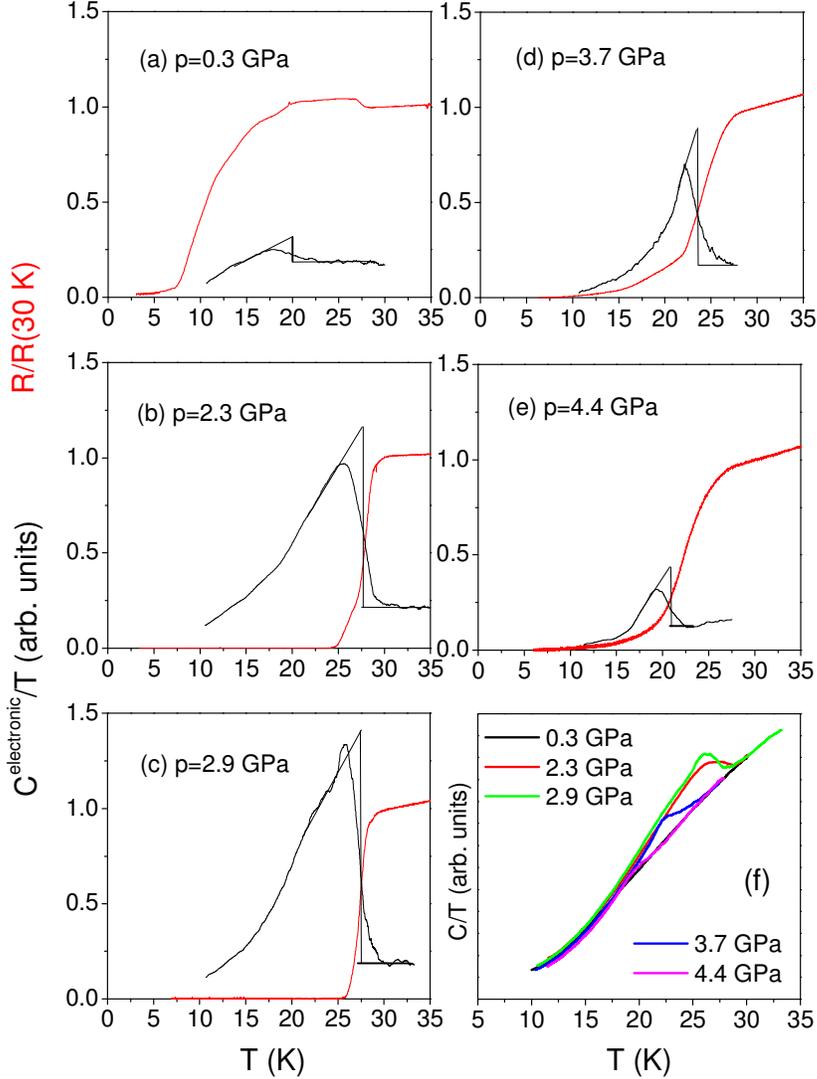

**Figure. 4** Approximate electronic specific heat in comparison with resistivity data (normalized at 30 K) of Ba(Fe$_{0.963}$Co$_{0.037}$)As$_2$ at various pressures up to 4.4 GPa. The lines help to estimate the size of the anomaly without broadening. The total heat capacity for all applied pressures before subtracting an approximated phonon background is shown in (f).

The midpoint of the jump coincides with the onset of the resistive transition. However, the resistance only gradually approaches zero resistance well below 5 K, which is in contrast to the maximum of the specific heat at ~18 K. Although specific heat clearly reveals a bulk nature of superconductivity below ~ 18 K, the pressure inhomogeneity at this low initial pressure causes finite resistance, either caused by strain-induced microcracks in the sample or due to scattering on twin boundaries of the orthorhombic phase [43]. The small step at 27 K may be a signature of the formation of such twin boundaries. The specific heat shows no anomaly at this temperature. Therefore, this process is only associated to a minor change in entropy.

Upon increasing pressure, the specific-heat jump grows rapidly in magnitude. The maximum $T_c$ is reached for the second pressure of 2.3 GPa (Figure 4b) with zero resistance at 25 K and an onset at ~30 K. At 2.3 GPa and 2.9 GPa the midpoint of the specific heat jump agrees perfectly with the midpoint of the resistive transition and also the transition widths are comparable. For higher pressures, $T_c$ drops towards lower temperature, indicating that the overdoped regime of the phase diagram is entered.

Starting from 3.7 GPa (Figure 4d), the anomaly transforms into a rather symmetric anomaly, which we attribute to broadening effects arising from our solid pressure medium. The size of the specific heat anomaly decreases now as function of pressure, which is mostly a consequence of the $T_c$ reduction on the overdoped side. At 4.4 GPa (Figure 4e) the superconducting transition is suppressed to 22 K. Upon comparing the specific-heat data with resistivity in this pressure range, one notices that the specific-heat transition extends over the full width of the more or less broadened resistive transition, although the resistance shows some tail in the low temperature regime, which may be a consequence of internal stress in the sample due to pressure gradients.

## IV. DISCUSSION

In Figure 5(a) we plot a critical temperature vs. pressure phase diagram as derived from our data. Our study differs from most of the previous high-pressure studies [1-5] in a sense that we do not start from the parent undoped Ba122 compound but from an underdoped Co-substituted sample, which is exactly at the borderline of superconductivity. We observe a transition temperature with a $T_c$ onset almost approaching 30 K at optimal pressure between 2 and 3 GPa. Our high-pressure specific-heat technique reveals large anomalies at $T_c$ over the whole pressure range which proves that at least a major volume fraction the sample becomes superconducting. Pressure-induced superconductivity with onset temperatures up to 29 K has been observed previously in undoped Ba122 [1-5]. However, our bulk thermodynamic method provides us additional information. Upon comparison of the two methods there is a clear difference in the determination of $T_c$. What agrees well is the onset $T_c$ in the underdoped and optimally-doped pressure range. In the highest pressure (4.4 GPa) the resistive onset remains at somewhat higher temperature, which may indicate the presence of filamentary superconductivity above ~25 K for which the specific heat is not sensitive. Furthermore, the temperatures where the resistivity reaches zero deviate strongly on both, the underdoped and the overdoped side of the phase diagram.

The size of the specific-heat anomaly $\Delta C$ at $T_c$ is closely related to the superconducting condensation energy $U_0$ [33]. While $\Delta C$ can be directly obtained from our data, the condensation energy $U_0$ needs to be derived by integration of the specific heat according to Eq. (1). The unknown phonon contribution causes certainly a significant imprecision, however the pressure-induced variation is nevertheless reflected in the so-derived data.

$$U_0 = \int_T^{T_c} \left[ C_s(T') - C_n(T') \right] dT' \qquad (1)$$

In Figure 5(b) the results are plotted. In order to compare our data of $\Delta C$ and $U_0$ qualitatively with ambient-pressure data [29], we used the formula suggested by Drotzinger et al. [31], which converts the effect of pressure into a corresponding variation of Co-content: $\Delta P/\Delta x \approx 1.275$ GPa / at.% Co. Being lack of an absolute value, we furthermore scaled our $\Delta C$ data at the lowest pressure to the corresponding ambient-pressure value. $U_0$ has been scaled to $\Delta C$ for comparison. The largest specific-heat anomaly is observed at optimal doping/pressure with the maximum value of $\Delta C$ exceeding the corresponding value at ambient pressure clearly. Both, $\Delta C$ (triangles) and $U_0$ (squares) show a similar trend with a sharp peak at 2.9 GPa. This indicates a strong pressure-induced increase in the condensation energy, qualitatively similar to the doping dependence [29], but of larger magnitude. The stronger pressure induced increase compared to ambient pressure data is furthermore directly visible upon comparing the specific heat data: the observed specific-heat anomaly at $T_c$ under optimally Co-doped conditions and ambient pressure represents ~10 % of the total specific heat. Under pressure, the anomalies represent up to 20 % of the total specific heat, while the optimal $T_c$ increases only from 25 K to 27.7 K (as defined from the midpoint of the specific heat jump). Note, that this value may even be underestimated as additional phonon contributions from the surrounding pressure medium likely enter the signal.

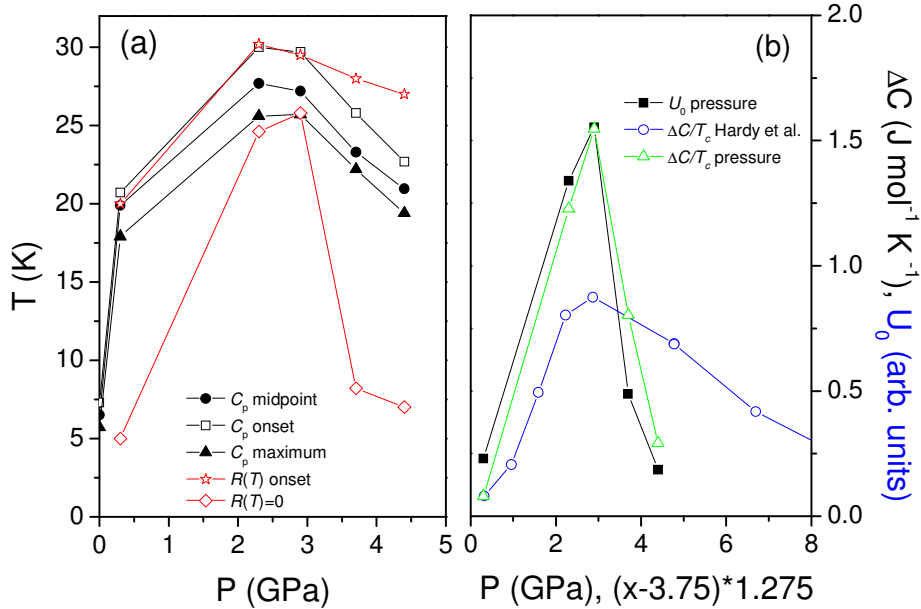

**Figure. 5** (**a**) High-pressure phase diagram of Ba(Fe$_{0.963}$Co$_{0.037}$)As$_2$ with the superconducting transition as obtained from specific heat and resistivity as function of applied pressure. In order to illustrate the width of the transition, we included the onset of the specific-heat transition, the midpoint of the jump-like specific-heat anomaly (inflection point) and the specific-heat maximum. For the resistivity, we plotted the onset of the resistive transition and the point where the resistance reaches approximately zero. (**b**) Plot of the pressure-induced variation of the superconducting condensation energy $U_0$ and the closely-related quantity $\Delta C$, as obtained from our pressure specific-heat data in comparison to $\Delta C$ data from samples with various Co contents $x$ at ambient pressure [29]. The $\Delta C$ pressure data has been scaled to the ambient pressure data at 0.3 GPa. For comparison of the Co concentration with the pressure scale the formula suggested by Drotzinger et al. [31] was used.

The data demonstrates that under pressure a much larger condensation energy and coupling strength is found than in optimally Co-doped samples. The maximum does not appear at the optimal pressure (2.3 GPa), but rather at 2.9 GPa. It is quite interesting that the pressure-induced variation of $U_0$ follows closely the trend of the zero-resistance line. The variation of $U_0$ therefore is likely linked to the homogeneity of the superconducting state. This agrees with the finding that a residual Sommerfeld constant term is found in ambient-pressure specific-heat data, which approaches a minimum value at optimal doping [29]. At a pressure of 2.9 GPa the superconducting state would therefore be most homogeneous. The observed clear difference in the optimal $T_c$ and in $U_0$ demonstrates that the application of pressure cannot be regarded as solely equivalent to Co-substitution. There must be an additional mechanism, which either strengthens the superconducting state under pressure (and thus increases the optimal $T_c$ and $U_0$); or which weakens the superconducting state at ambient pressure upon Co doping. The superconducting phase under pressure appears furthermore much narrower in the overdoped pressure regime.

A mechanism that could increase $T_c$ and the condensation energy under pressure may be found in the pressure-induced changes in the crystalline structure. However, a more likely and natural explanation is that the superconducting state in optimally Co-doped samples at ambient pressure is weakened. In contrast to $K^+$ doping on the Ba site, which causes somewhat higher optimal $T_c$ values, Co replaces Fe atoms directly in the, for superconductivity critical, FeAs layers. It is already surprising, that the superconductivity in Ba122 is so robust against chemical substitution of $Fe^{2+}$ with $Co^{3+}$, which contrasts to the cuprates, where substitution of only a small percentage of Cu atoms immediately destroys superconductivity [44]. The effect of various dopants on impurity scattering and pair-breaking in the pnictides has been studied e.g. for Mn and Zn [45] and for trivalent charge state rare-earth ions such as La, Ce, Pr and Nd [46]. It has been demonstrated that magnetic scattering plays a significant role on the maximum $T_c$ value, whereas impurity scattering (e.g. in Zn-doped samples) has a much smaller pair breaking effect. The larger values of $T_c$ and the condensation energy of our underdoped sample under optimal pressure suggest that Co also acts as a weak pair breaker, since it contains far less Co than optimally-doped samples at ambient pressure. It has been shown that $\Delta C$ varies in various Fe-based superconductors as a function of $T_c^3$ [47]. This scaling differs dramatically from classical BCS superconductors, for which it is expected that $\Delta C$ depends linearly on $T$. The reason for this unusual relation in the Fe-based superconductors is strongly debated, but has been explained by the presence of a quantum critical point within the superconducting dome [48], or linked to pair breaking effects [49,50]. Our pressure $\Delta C$ data increases even faster with increasing $T_c$ values in the underdoped regime than the ambient pressure $T_c^3$ dependence. This may further confirm that under pressure the pair breaking is reduced in accordance with the pair breaking scenario proposed by V. G. Kogan [49,50].

## V. CONCLUSIONS

To conclude, we were able to trace the high-pressure phase diagram, starting from a strongly underdoped single crystal of $Ba(Fe_{0.963}Co_{0.037})As_2$, using specific heat as a bulk thermodynamic probe in combination with electrical resistivity. The availability of this bulk thermodynamic quantity proved to be a powerful tool to investigate the phase diagram of Fe-based superconductors under pressure. In contrast to the resistivity, which may be influenced by filamentary superconductivity and scattering on structural domain walls or orthorhombic twin boundaries [43], the specific-heat jumps are comparably sharp and indicate the true bulk $T_c$. A strong pressure-induced increase of the maximum $T_c$ and the superconducting condensation energy is observed, which demonstrates that using pressure as a control parameter is not exactly equivalent to Co-substitution. A magnetic Cooper-pair-breaking effect of Co-doping is discussed as the most likely explanation of the lower $T_c$ values of Co-doped samples under ambient pressure conditions.

**Acknowledgments**
R. L. thanks U. Lampe for technical support D. Jaccard for sharing his high-pressure secrets with him. This research is inspired by an idea of A. Junod. This work was supported by the Research Grants Council of Hong Kong, Grant No. HKUST 603010 and SEG_HKUST03 and has been partially funded through the DFG priority program (SPP1458) "High-Temperature Superconductivity in Iron Pnictides" of the Deutsche Forschungsgemeinschaft (DFG).